\documentclass[11pt,leqno]{article} 
\usepackage{epsfig}
\usepackage{isolatin1}
\usepackage{a4}
\usepackage{amsmath, amsfonts, amssymb, latexsym} 
\usepackage{myth}

\newcommand{\du}{{\protect\rule{0mm}{1mm}}\ifmmode\mathrm{d}
  \else\rmfamily{d} \fi}
\newcommand{\G}{\ifmmode {\mathcal G} \else $\mathcal G$\relax\fi}
\newcommand{\Gp}{\ifmmode {\mathcal G'} \else $\mathcal G'$\relax\fi}
\newcommand{\U}{\ensuremath{\mathcal U}}
\newcommand{\N}{\ensuremath{\mathbb N}}
\newcommand{\R}{\ensuremath{\mathbb R}}
\def\beq #1\eeq {\begin{align} #1 \end{align}}
\newcommand{\equi}{{\protect\rule{0mm}{0mm}}\ifmmode%
  \quad \Longleftrightarrow \quad \else $\Leftrightarrow$ \fi}
\newcommand{\thus}{{\protect\rule{0mm}{0mm}}\ifmmode%
  \quad \Longrightarrow \quad \else $\Rightarrow$ \fi}
\newcommand{\punkt}{~.}
\newcommand{\komma}{~,}
\newcommand{\fig}[3]{\begin{figure}[#3]
\centerline{\psfig{file=#2.eps,silent=}} 
\caption{\label{fig#2}#1}
\end{figure}}

\newcommand{\mbb}{\mathbb}

\begin{document}
\allowdisplaybreaks
\hfuzz=0.3cm

\thispagestyle{empty}
\begin{center}
\vspace*{1.0cm}

{\LARGE{\bf Pregeometric Concepts on Graphs and \\ Cellular
    Networks as Possible Models of \\[0.3cm] Space-Time at the
    Planck-Scale}\footnote{Invited paper to appear in the special issue
of the Journal of Chaos, Solitons and Fractals on: ``Superstrings, M, F,
S \dots Theory'' (M.\ S.\ El Naschie, C.\ Castro, Editors)}}

\vskip 1.5cm

{\large {\bf Thomas Nowotny\quad Manfred Requardt }} 

\vskip 0.5 cm 

Institut f\"ur Theoretische Physik \\ 
Universit\"at G\"ottingen \\ 
Bunsenstrasse 9 \\ 
37073 G\"ottingen \quad Germany

\end{center}

\vspace{1cm}

\begin{abstract}
  Starting from the working hypothesis that both physics and the
  corresponding mathematics have to be described by means of discrete
  concepts on the Planck-scale, one of the many problems one has to
  face is to find the discrete protoforms of the building blocks of
  continuum physics and mathematics. In the following we embark on
  developing such concepts for irregular structures like (large)
  graphs or networks which are intended to emulate (some of) the
  generic properties of the presumed combinatorial substratum from
  which continuum physics is assumed to emerge as a coarse grained and
  secondary model theory. We briefly indicate how various concepts of
  discrete (functional) analysis and geometry can be natu\-ral\-ly
  constructed within this framework, leaving a larger portion of the
  paper to the systematic developement of dimensional concepts and
  their properties, which may have a possible bearing on various
  branches of modern physics beyond quantum gravity.
\end{abstract} \newpage \setcounter{page}{1}

\section{Introduction} \label{sec1}
There exists a certain suspicion in parts of the scientific community
that nature may be discrete or rather ``behaves discretely'' on the
Planck scale. But even if one is willing to agree with this ``working
philosophy'', it is far from being evident what this vague metaphor
actually has to mean or how it should be implemented into a
concrete and systematic inquiry concerning physics and mathematics in the
Planck regime.

There are basically two overall attitudes as to ``discreteness on the
Planck scale'', the one comprising approaches which start (to a
greater or lesser degree) from continuum concepts (or more
specifically: concepts being more or less openly inspired by them) and
then try to detect or create modes of ``discrete behavior'' on very
fine scales, typically by imposing quantum theory in full or in parts
upon the model system or framework under discussion.

There are prominent and very promising candidates in this class like
e.g. {\em `string theory'} or {\em `loop quantum gravity'}. Somewhat
intermediate is a more recent version (or rather: aspect) of the
latter approach, its {\em `polymer'} respectively {\em `spin
  network'} variants. As these approaches are widely known and will
probably get their fair share in this volume anyhow, we refrain from
citing from the vast corresponding literature. We only recommend, as
more recent reviews to the latter approach, containing some cursory
remarks about the former together with a host of references, \cite{1} 
and \cite{2} and, as a beautiful introduction to the whole field,
\cite{3}.

On the other hand one could adopt an even more speculative and radical
attitude and approach the Planck regime from more or less the opposite
direction by developing a framework almost from scratch which has
``discreteness'' already built in into its very building blocks and
then try to reconstruct, working, so to speak, ``bottom up'', all the
continuum concepts of ordinary space-time physics as sort of
{\em `collective quantities'} like e.g. {\em `collective excitations'} via
the cooperation of many microscopic (discrete) degrees of freedom. If
one is very bold one could even entertain the idea that the quantum
phenomena as such are perhaps not the eternal and irreducible
principles as they are still viewed today by the majority of
physicists but, to the contrary, may emerge as derived and secondary
concepts, together with gravitation, from a more primordial truly
discrete and {\em `combinatorial'} theory. 

It goes without saying that such a radical approach is beyond the
reach of direct experimental verification in the strict sense for the
foreseeable future (as is the case with the other frameworks mentioned
above). As a substitute one rather has to rely on inner theoretical
criteria, among other things the capability to generate the hierarchy
of complex patterns we are observing in nature or in our present day
{\em `effective theories'} of the various regimes many orders of
magnitude away from the Planck scale while introducing as few simple
and elementary assumptions as possible. More specifically, one would
like such a framework to provide clues how the continuum concepts of
ordinary space-time physics/mathematics may emerge in a natural
manner from their respective {\em `discrete protoforms'}.

Another more aesthetic criterion would be a kind of natural
convergence of the different approaches towards a common substructure
which is discrete in a really primordial way. Indications for such a 
convergence can, in our view, be detected in the various lines of
research going on presently. {\em `spin networks'} or {\em `polymer
  states'} are e.g. cases in point where modes of discreteness do
emerge from a at first glance continuous environment. Furthermore it
may well be that {\em `string field theory'} will turn out to live on
a more discrete and singular substratum than presently suspected (some
speculative remarks pointing in this direction can e.g. be found at
the end of \cite{4}), a catchword being {\em `fractal geometry'}. A
brief but illuminating analysis concerning such a possible convergence
in the future towards a decidedly discrete and {\em `combinatorial'}
common limit has been given in section 8 of \cite{3}. A bundle of
related ideas with which we sympathize is developed by Nottale (see
e.g. \cite{12}).

In the following we will try to give a very brief survey over our
personal variant of Planck scale physics and mathematics, which has a
pronouncedly combinatorial flavor (employing, among other things, tools from
{\em `algebraic combinatorics'} and {\em `(random) graph
  theory'}). Interesting enough, there exist also close ties to
{\em `noncommutative geometry'} (a general source being \cite{5}).

\section{The Cellular Network Environment}
In this section we will sketch the type of model systems on which the
following analysis will be based. As already said in the introduction
we will start from a rather primordial level, trying to make no
allusions whatsoever to continuum concepts. We will then show how
protoforms of ideas and notions, playing a key role in ordinary
continuum physics/mathematics emerge in a relatively natural and
unforced way from this framework. Cases in point are e.g. concepts
like {\em `dimension'}, {\em `differential structure'}, the idea of
{\em `physical points'} (being endowed with an internal structure),
the web of which establishes the substratum of macroscopic space-time,
and other geometrical/topological notions. The framework turns out to
be even rich enough to support a full fledged kind of
{\em `discrete functional analysis'}, comprising e.g. {\em `Laplace-,
  Dirac operators'} and all that. It is perhaps particularly
noteworthy that an advanced structure like {\em `Connes' spectral
  triple'} shows up in a very natural way within this context.

Besides the reconstruction of basic concepts of continuum
physics/ mathematics another goal is to describe the micro dynamics
going on in this discrete substratum over (in) which macroscopic
space-time is floating as a kind of coarse grained
{\em `superstructure'}, the formation of {\em `physical points'} and
their mutual entanglement, yielding the kind of
{\em `near-/far-order'} or {\em `causal structure'} we are used to from
continuum space-time.

To this end we view this substratum as, what we like to call, a
{\em `cellular network'}, consisting of {\em `nodes'} and
{\em `bonds'}. The nodes are assumed to represent certain elementary
modules (cells or ``monads'') having a discrete, usually simple,
internal state structure, the bonds modeling elementary direct
interactions among the nodes. As an important ingredient, these bonds
are dynamical insofar as they are capable to be in a (typically limited)
number of {\em 'bond states'}, thus implementing the varying strength
of the mutual interactions among the cells.

It is a further characteristic of our model class that these
interactions are not only allowed to vary in strength but, a fortiori,
can be switched off or on, depending on the state of their local 
environment. In other words, bonds can be, stated in physical terms,
created or annihilated in the course of network evolution, which
(hopefully) enables the system to undergo {\em `geometric phase
  transitions'} being accompanied by an {\em `unfolding'} and
{\em `pattern formation'}, starting e.g. from a less structured
chaotic initial phase. To put it briefly: in contrast to, say,
{\em `cellular automata'}, which are relatively rigid and regular in
their wiring and geometric structure (in particular: with the bonds
typically being non-dynamical), our cellular networks do not carry such
a rigid overall order as an external constraint (e.g. a regular
lattice structure); their ``wiring'' is dynamical and thus behaves randomly
to some extent. The clue is that order and modes of regularity are
hoped to emerge via a process of {\em `self-organization'}.
\begin{definition}[(Class of Cellular Networks)] \hfill  
\begin{enumerate}
\item ``Geometrically'' our networks represent at each fixed
{\em `clock time'} a {\em `labeled graph'}, i.e. they consist of nodes
\{$n_i$\} and bonds \{$b_{ik}$\}, with the bond $b_{ik}$ connecting
the nodes (cells) $n_i$, $n_k$. We assume that the graph has neither
elementary loops nor multi-bonds, that is, only nodes with $i\neq k$
are connected by at most one bond.
\item At each site $n_i$ we have a local node state $s_i\in
q\cdot\mbb{Z}$ with $q$, for the time being, a certain not further
specified  elementary quantum. The bond variables $J_{ik}$, attached
to $b_{ik}$, are in the most simplest cases assumed to be two- or
three-valued, i.e. $J_{ik}\in \{\pm 1\}\quad\text{or}\quad J_{ik}\in \{\pm
  1,0\}$
\end{enumerate}
\end{definition}
\begin{remark} \hfill 
\begin{enumerate}
\item In the proper graph context the notions {\em `vertex'} and 
{\em 'edge'} are perhaps more common (see e.g. \cite{6}). As to some
further concepts used in graph theory see below.
\item These are, in some sense, the most simple choices one can
make. It is an easy matter to employ instead more complicated internal
state spaces like, say, groups, manifolds etc. One could in particular
replace $\mbb{Z}$ by one of its subgroups or impose suitable  boundary 
conditions.
\item In the following section we will give the bonds $b_{ik}$ an 
{\em 'orientation'}, i.e. (understood in an precise
algebraic/geometric sense) $b_{ik}=-b_{ki}$.  
This implies the compatibility conditions $J_{ik}=-J_{ki}$.
\end{enumerate}
\end{remark}

In a next step we have to impose a dynamical law on our model
network. In doing this we are of course inspired by {\em `cellular
  automaton laws'} (see e.g. \cite{7}). The main difference is however
that in our context also the bonds are dynamical degrees of freedom
and that, a fortiori, they can become dead or alive (active or
inactive), so that the whole net is capable of performing drastic 
topological/geometrical changes in the course of clock time.

A particular type of a dynamical '{\it local law}' is now introduced
as follows: We assume that all the nodes/bonds at '{\it (clock) time}'
$t+\tau$, $\tau$ an elementary clock time step, are updated according
to a certain local rule which relates for each given node $n_i$ and
bond $b_{ik}$ their respective states at time $t+\tau$ with the states
of the nodes/bonds of a certain fixed
local neighborhood at time $t$.

It is important that, generically, such a law does not lead to a
reversible time evolution, i.e. there will typically exist attractors
in total phase space (the overall configuration space of
the node and bond states).

A crucial ingredient of our network
laws is what we would like to call a '{\it hysteresis interval}'. We
will assume that our network starts from a densely entangled '{\it
  initial phase}' $QX_0$, in which practically every pair of nodes is
on average connected by an '{\it active}' bond, i.e. $J_{ik}=\pm1$.
Our dynamical law will have a built-in mechanism which switches bonds
off (more properly: sets $J_{ik}=0$) if local fluctuations among the node
states become too large. Then there is hope that this mechanism
may trigger an '{\it unfolding phase transition}', starting from a
local seed of spontaneous large fluctuations towards a new phase (an
attractor) carrying a certain '{\it super structure}', which we would
like to relate to the hidden discrete substratum of space-time
(points).

One example of such a law is given in the following definition.

\begin{definition}[(Local Law)]
At each clock time step a certain '{\it quantum}' $q$
is transported between, say, the nodes $n_i$, $n_k$ such that 
\begin{align}s_i(t+\tau)-s_i(t)=q\cdot\sum_k
  J_{ki}(t)\end{align}
(i.e. if $J_{ki}=+1 $ a quantum $q$ flows from $n_k$ to $n_i$ etc.)\\
The second part of the law describes the back reaction on the bonds
(and is, typically, more subtle). This is the place where the
so-called '{\it hysteresis interval}' enters the stage. We assume the
existence of two '{\it critical parameters}'
$0\leq\lambda_1\leq\lambda_2$ with:
\begin{align}J_{ik}(t+\tau)=0\quad\mbox{if}\quad
  |s_i(t)-s_k(t)|=:|s_{ik}(t)|>\lambda_2\end{align}
\begin{align}J_{ik}(t+\tau)=\pm1\quad\mbox{if}\quad 0<\pm
  s_{ik}(t)<\lambda_1\end{align}
with the special proviso that
\begin{align}J_{ik}(t+\tau)=J_{ik}(t)\quad\mbox{if}\quad s_{ik}(t)=0
\end{align}
On the other side
\begin{align}J_{ik}(t+\tau)= \left\{\begin{array}{ll} 
\pm1 & \quad J_{ik}(t)\neq 0 \\
0    & \quad J_{ik}(t)=0
\end{array} \right. \quad\mbox{if}\quad
\lambda_1\leq\pm
  s_{ik}(t)\leq\lambda_2 \punkt
\end{align}
In other words, bonds are switched off if local spatial charge
fluctuations are too large, switched on again if they are too
small, their orientation following the sign of local charge
differences, or remain inactive.
\end{definition}
\begin{remark} \hfill 
\begin{enumerate}
\item The reason why we do not choose the ''current'' $q\cdot
J_{ik}$ proportional to the ''voltage difference'' $(s_i-s_k)$ as e.g.
in Ohm's law is that we favor a non-linear (!) network which is
capable of self-excitation and self-organization rather than
self-regulation around a relatively uninteresting equilibrium state!
The balance between dissipation and amplification of spontaneous
fluctuations has
however to be carefully chosen (``complexity at the edge of chaos'')
\item We presently have emulated these local network laws on a computer.
It is not yet clear whether this simple network law already does
what we expect. In any case, it is fascinating to observe the
enormous capability of such intelligent networks to find attractors
very rapidly, given the enormous accessible phase space
\item In the above class of laws a direct bond-bond-interaction is not
yet implemented. We are prepared to incorporate such a contribution in
a next step if it turns out to be necessary. In any case there are not
so many ways to do this in a sensible way. Stated differently, the class
of possible physically sensible interactions is
perhaps not so numerous.
\item Note that -- in contrast to e.g. Euclidean lattice field theory --
the so-called {\it `clock time'} $t$ is, for the time being, not
standing on the same footing as potential ``coordinates'' in the
network (e.g.  curves of nodes/bonds). Anyhow We suppose that so-called
{\it `physical time'} will emerge as sort of a secondary collective
variable in the network, i.e. being different from the clock time
(while being of course functionally related to it). \label{no4}
\end{enumerate}
\end{remark}

In our view \ref{no4} is consistent with the spirit of relativity. What
Einstein was really teaching us is that there is a (dynamical)
interdependence between what we experience as space respectively time, not
that they are absolutely identical! In any case the assumption of an
overall clock time is at the moment only made just for convenience in
order to make the model system not too complicated. If our
understanding of the complex behavior of the network dynamics
increases, this assumption may be weakened in favor of a possibly
local and/or dynamical clock frequency. A similar attitude should be
adopted concerning concepts like {\it `Lorentz-(In)Covariance'} which
we also consider as {\it `emergent'} properties. It is needless to say that
it is of tantamount importance to understand the way how these
patterns do emerge from the relatively chaotic background which will
be attempted in future work.

As can be seen from the definition of the cellular network, a full
scale investigation of its behavior
separates quite naturally into two parts of both a different
mathematical and physical nature. The first one comprises its more
geometric/algebraic content in form of large static graphs and their intricate
structure (at, say, arbitrary but fixed clock time), thus neglecting
the details of the internal states of bonds and nodes, the other one
conveys a more dynamical flavor, i.e. analyzing
topological/geometrical change and pattern formation in the course of
clock time.  

Due to lack of space we cannot treat all these different aspects in
any detail, as their proper discussion would require, among other
things, the development of a fair amount of relatively advanced
(discrete) mathematics. Therefore we prefer to make only a couple of
provisional remarks as to some of the necessary building blocks of our
framework in the following section, referring to our other papers for
more details, and using the remaining space for a discussion of one
single aspect in slightly more depth. This aspect is the development of a
suitable concept of {\em `dimension'} on graphs and similar erratic structures.

\section{A brief Discussion of various Geometric and Topological Concepts
  on Graphs and Networks}
In a first step we would like to have something like a {\em `discrete
differential calculus'} and {\em `functional analysis on graphs'} as kind of
protoforms of the corresponding continuum concepts. We recently have
shown, that this is in fact possible and leads to quite interesting
mathematical structures. As to differential calculus and related
aspects see the corresponding sections in \cite{8} and references
given there as well as \cite{Di} for a complimentary but slightly
different approach. Functional
analysis on graphs is developed in \cite{9}. As to some other aspects
of discrete noncommutative geometry refer to the work of Sorkin and
Balachandran et al \cite{So}, \cite{Ba} .

The starting point of our approach is the introduction of a
differential calculus on {\em `node-'} and {\em `bond functions'} as
elements of the {\em 'node-'} and {\em `bond space'} of a given fixed
graph $G$. To put it briefly, one can define the derivative of
elementary node functions $n_i$ (the function with the value one on
the node $n_i$, zero elsewhere) and extend it by linearity to general
node functions $f:=\sum f_i\cdot n_i$, the sum taken over the set of
nodes. It is then an easy matter to define e.g. $L^p$-spaces etc. The
crucial point is that the derivative $dn_i$ is represented as the sum
over the oriented bonds being incident with $n_i$, i.e.
\begin{align} dn_i:=\sum_k b_{ki}\end{align}
(where the $n_i$'s and $b_{ik}$ are viewed to generate complex vector
spaces). In other words, $d$ maps the node space into the bond space.

It turns out to make sense to introduce in addition to the oriented
bonds, $b_{ik}$, {\em `directed bonds'}, $d_{ik}$, having a fixed
direction, pointing from node $n_i$ to node $n_k$ with
\begin{align} b_{ik}:=d_{ik}-d_{ki}\end{align}
This yields:
\begin{align}df=\sum_{ik}(f_k-f_i)d_{ik}\end{align}
If one develops this discrete calculus further quite a few interesting
aspects will emerge with links to various areas of modern mathematics,
catchwords being: {\em `non-commutative geometry'}, {\em `modules'},
{\em `groupoids'} etc. (\cite{8}). 

It is also possible to develop something like discrete functional
analysis on graphs (\cite{9}). Among other things a {\em `graph
  Laplacian'} does exist which turns out to be intimately related with
the {\em `adjacency matrix'} of graph theory.
\begin{align}
-\Delta f:=-\sum_i(\sum_k(f_k-f_i))n_i \quad \text{and} \quad
-\Delta=V-A \komma
\end{align}
with $A$ the adjacency matrix of a graph (entries
$a_{ik}=1$ or $0$ depending on whether the nodes $n_i$ and
$n_k$ are connected by a bond or not). $V$ is the {\em `vertex degree
  matrix'}, its diagonal elements $v_{ii}$ counting the number of
bonds being incident with $n_i$ (for more details see the literature
cited in \cite{9}). It is particularly noteworthy that $-\Delta$ or
$A$ encode many geometric/combinatorial graph properties being of
general interest.

\section{The Random Graph Aspect of the Dynamical Network and the
  Notion of Physical Points}

We again have to be very brief; for more details we refer to
\cite{10}, which represents, however, given the rapid development of
this new field, only a preliminary draft. A more up to date version is
forthcoming. The underlying idea is the following: Instead of studying
the extremely complicated network dynamics in full, it is tempting to
try to catch only its generic qualitative behavior. Following this
idea a statistical approach suggests itself. One could e.g. assume
that the network dynamics is sufficiently random so that {\it `graph
properties'} can be modeled as {\it `random functions'} over a
certain probability space. In a completely different spirit similar
ideas have been developed quite some time ago by Erd\H{o}s and Renyi and
more recently by e.g. Bollob\'as (see \cite{11}). It was then an important
observation that many graph properties have a so-called {\it `threshold
  function'}, which is very reminiscent of a {\it `phase transition
  line'} in statistical physics.

We already remarked above that we are particularly interested in the
possibility of geometric phase transitions in our network. Our hope is
that something like a protoform of space-time may emerge as kind of a
superstructure in the network. The elementary building blocks of this
fabric, which we like to call `{\it physical points}' we expect to be
made up of densely wired `{\it sub-clusters}' of nodes/bonds. These, on
their side, are then assumed to establish a kind of near-/far-order
in the network, thus generating something like a causal structure. One
possibility to associate what we like to call `{\it physical points}' 
with a certain class of subgraphs is it to define them as the 
`{\it maximally connected subgraphs}' (i.e. subgraphs which are
maximal simplices) or what graph theorist call a
`{\it clique}'. Such cliques can be constructed or found in an
algorithmic way starting from an arbitrary node (\cite{10}).

\section{More Graph Theoretical Definitions} \label{sec2}
In this section we give some more definitions of ordinary graph theory
to discuss one of the aspects mentioned above, the dimensional concept
on graphs, in more detail in the next section. Most of the notions are
well known in graph theory but we nevertheless want to repeat them to
avoid any confusion concerning the exact definitions.

We already introduced the undirected simple graph as the geometric
aspect of a cellular network. 
In the following $\G=(N,B)$ will always be an undirected simple
graph. We also need the notion of the degree of a node $n_i \in N$.
\begin{definition}[(Degree)] \label{def2}
The {\em degree} of a node $n_i \in N$ is the number of bonds incident with
it, i.e. the number of bonds which have $n_i$ at one end. We 
count $b_{ik}$ and $b_{ki}$ only once as we interpret them as the same
bond.
\end{definition}
We assume the node degree of any node $n_i \in N$ of the graphs under
consideration to be finite.
The next step is to define a metric structure on \G. To this end we
need to define paths in \G ~and their length.
\begin{definition}[(Path)] \label{def3}
A {\em path} $\gamma$ of length $l$ in \G ~is an ordered $(l+1)$ tuple
of nodes $n_i \in N$, $i \in I$, $I=\{0,\dots,l\}$ with the properties
$n_{i+1} \neq n_{i}$ and $b_{i \, i+1} \in B$.
\end{definition}
A single node $n_i \in N$ is a path of length $0$. This definition
encodes the obvious idea of a path in \G ~allowing multiple
transversals of nodes or bonds.  Jumps across non-existent bonds and
stays at a single node are not allowed. Sometimes this notion of a
path is also called a {\em bond sequence}.

We will call a path with the property that all $n_i \in \gamma$ are
pairwise different a {\em simple path}.

The concept of paths on \G ~now leads to a natural definition for the
distance of two nodes $n_i$ and $n_j \in N$, namely the length of the
shortest path connecting $n_i$ and $n_j$. 
\begin{definition}[(Metric)] \label{def4}
Let $l(\gamma)$ denote the length of $\gamma$. A {\em metric} $d$ on \G ~is
\begin{align}
d(n_i,n_j) := \left\{ \begin{array}{lc} {\min}\{l(\gamma): n_i, n_j \in
    \gamma \} & \mbox{if such $\gamma$ exist} \\ \infty &
    \mbox{otherwise.}
\end{array} \right.
\end{align}
\end{definition}
That this actually defines a metric is easily established. Finally we
need the notion of neighborhoods which follows canonically from the
metric.
\begin{definition}[(Neighborhood)] \label{def5}
Let $n_i \in N$ be an arbitrary node in \G. An $n$- neighborhood of
$n_i$ is the set $\U_n(n_i):=\{n_j \in N: d(n_i, n_j) \leq n\}$.
\end{definition}
\begin{remark}
The topology generated by the $n$-neighborhoods is the discrete
topology as should be expected from the construction and the
discreteness of graphs.
\end{remark}
We will denote the {\em surface} or {\em boundary} of the neighborhood
$\U_n(n_i)$ as $\partial\U_n(n_i)
:= \U_n(n_i) ~\backslash ~\U_{n-1}(n_i)$, $\partial \U_0(n_i)= \{n_i\}$
and the cardinality of $\U_n(n_i)$ and $\partial \U_n(n_i)$ as
$|\U_n(n_i)|$ and $|\partial \U_n(n_i)|$ respectively.

\section{Dimensions of Graphs and Networks} \label{sec3}
\begin{definition}[(Internal Scaling Dimension)] \label{def6}
  Let $x \in N$ be an arbitrary node of ~\G. Consider the sequence of
  real numbers $D_n(x):= \frac{\ln|\U_n(x)|}{\ln(n)}$. We say
  $\underline{D}_S(x):= \liminf_{n \rightarrow \infty} D_n(x)$ is the
  {\em lower} and $\overline{D}_S(x):= \limsup_{n \rightarrow \infty}
  D_n(x)$ the {\em upper internal scaling dimension} of \G ~{\em
  starting from $x$}. If $\underline{D}_S(x)= \overline{D}_S(x)=: D_S(x)$ we
  say {\G ~has internal scaling dimension $D_S(x)$ starting from
  $x$}. Finally, if $D_S(x)= D_S$ $\forall x$, we simply say \G ~has
  {\em internal scaling dimension $D_S$}.
\end{definition}
A second notion of dimension we want to introduce is the {\em connectivity
dimension} which is based on the surfaces of neighborhoods 
$\partial\U_n(n_i)$ rather than on the whole neighborhoods $\U_n(n_i)$.
\begin{definition}[(Connectivity Dimension)] \label{def7}
Let $x \in N$ again be an arbitrary node of \G. We set $\tilde{D}_n(x)
:= \frac{\ln|\partial \U_n(x)|}{\ln(n)} +1$ and 
$\underline{D}_C(x) := \liminf_{n \rightarrow \infty} \tilde{D}_n(x)$ 
as the {\em lower} and $\overline{D}_C(x) := \limsup_{n \rightarrow 
\infty} \tilde{D}_n(x)$ as the {\em upper connectivity dimension}. 
If lower and upper dimension
coincide, we say \G ~has {\em connectivity dimension} $D_C(x) := 
\overline{D}_C(x) = \underline{D}_C(x)$ {\em starting from}
$x$. If $D_C(x) = D_C$ for all $x \in N$ we call $D_C$
simply the {\em connectivity dimension} of \G.
\end{definition}
One could easily think that both definitions are equivalent.
This is however not the case as one definition is stronger
than the other. We will discuss this in \ref{sec3.2}.

The internal scaling dimension is rather a mathematical concept
and is related to well known dimensional concepts in fractal 
geometry as we will see in \ref{sec4.2}. The connectivity dimension
on the other hand seems to be a more physical concept as it measures
more precisely how the graph is connected and thus how nodes can
influence each other.

\subsection{Basic Properties of the Internal Scaling Dimension}
\label{sec3.1} 
The first lemma gives us a criterion for the uniform
convergence of $\underline{D}_S(x)$ or $\overline{D}_S(x)$ to some common 
$\underline{D}_S$ or $\overline{D}_S$ for all nodes $x$ in \G.
\begin{lemma} \label{lem2}
Let $x$,$y \in N$ be two arbitrary nodes in \G ~with $d(x,y) < \infty$. 
Then $\underline{D}_S(y)=\underline{D}_S(x)$ and $\overline{D}_S(y)=
\overline{D}_S(x)$. 
\end{lemma}
\begin{proof}
Let $a:=d(x,y)$ be the distance of the nodes $x$ and $y$. We have
\begin{align}
& \U_{n-a}(y) \subseteq \U_n(x) \subseteq \U_{n+a}(y) \\ 
\thus & \frac{\ln|\U_{n-a}(y)|}{\ln(n)} \leq \frac{\ln|\U_n(x)|}{\ln(n)} 
\leq \frac{\ln|\U_{n+a}(y)|}{\ln(n)} \label{eqn5} \\
\thus & \frac{\ln|\U_{n-a}(y)|}{\ln(n-a) + \ln\big(\frac{n}{n-a}\big)} \leq 
\frac{\ln|\U_n(x)|}{\ln(n)}
\leq \frac{\ln|\U_{n+a}(y)|}{\ln(n+a)-\ln\big(\frac{n+a}{n}\big)} \\
\thus & \underline{D}_S(x) = \liminf_{n \rightarrow \infty} \frac{\ln|\U_n(x)|}
{\ln(n)} = \liminf_{n \rightarrow \infty} \frac{\ln|\U_n(y)|}{\ln(n)} 
= \underline{D}_S(y) \punkt
\end{align}
Similarly we get $\overline{D}_S(x)= \overline{D}_S(y)$.
\end{proof}

Another rather technical lemma provides us with a convenient method to
calculate the dimension of certain graphs, e.g. the self-similar or 
hierarchical graphs we construct in \ref{sec4.2}. It shows that under one
technical assumption the convergence of a subsequence of $D_n(x)$ is
sufficient for the convergence of $D_n(x)$ itself. 
\begin{lemma} \label{lem3}
Let $x \in N$ be an arbitrary node of \G ~and let $(|\U_{n_k}(x)|)_{k
  \in \N}$ be a subsequence of $(|\U_n(x)|)_{n \in \N}$. 
  There may exist a number $1 >
  c > 0$ such that $\frac{n_k}{n_{k+1}} \geq c$ holds for 
all $k \geq K \in \N$. 
Then $\liminf_{k \rightarrow \infty}
\frac{\ln|\U_{n_k}(x)|}{ln(n_k)} = \liminf_{n \rightarrow \infty} D_{n}(x)=
\underline{D}_S(x)$ 
and similar for $\overline{D}_S(x)$.
\end{lemma}
\begin{proof}
  The sequence of the neighborhood sizes $|\U_n(x)|$ is monotone such
  that $|\U_{n_k}(x)| \leq |\U_n(x)| \leq |\U_{n_{k+1}}(x)|$ for $n_k
  \leq n \leq n_{k+1}$. A short calculation yields
\begin{align}
\frac{\ln|\U_{n_k}(x)|}{\ln(n_k) + \ln(\frac{1}{c})}
\leq \frac{\ln|\U_n(x)|}{\ln(n)}
\leq \frac{\ln|\U_{n_{k+1}}(x)|}{\ln(n_{k+1}) + \ln(c)} \komma
\end{align}
which implies the conjecture.
\end{proof}
This result is well known in the context of calculation schemes for 
dimensions in fractal geometry, see e.g. \cite{6a}.

Naturally one also may ask how the internal scaling dimension
behaves under insertion of bonds into \G. We were able to show that it
is pretty much stable under any local changes. We state this in the
following lemma.
\begin{lemma} \label{lem4}
Let $k \in \N$ be
a positive natural number and $x \in N$ a node in \G. 
Insertion of bonds between arbitrary many pairs of nodes ($y$, $z$)
obeying the relation $d(y,z) \leq k$ does not change  
$\underline{D}_S(x)$ or $\overline{D}_S(x)$.
\end{lemma}
\begin{proof}
We denote the new graph built by insertion of new bonds into \G ~as \Gp
~and accordingly the neighborhoods in \Gp ~as $\U'_n(\cdot)$. Being a node
in \G, $x$ is also a node in \Gp. The restriction on the choice of
additional bonds in \Gp ~implies that even if we connect every node $y \in N$
with every node in $\U_k(y)$, which is the maximum we are allowed to do,
we still can't get beyond $\U_n(x)$ with less or equal 
$\lfloor \frac{n}{k} \rfloor$\footnote{The floor-symbol, $\lfloor x
  \rfloor$, denotes the largest integer below $x$, see
  e.g. \cite{Graham}} steps,
\begin{align}
  & \U_{\lfloor\frac{n}{k}\rfloor}(x) \subseteq 
  \U'_{\lfloor\frac{n}{k}\rfloor}(x) \subseteq \U_n(x) \punkt 
\end{align}
A short calculation yields the equality $\liminf_{n \rightarrow
  \infty} \frac{\ln|\U'_n(x)|}{\ln(n)} = \liminf_{n \rightarrow
  \infty} \frac{\ln|U_n(x)|}{\ln(n)}$. The same holds for $\limsup$.
\end{proof}
\begin{remark}
  Obviously the insertion of a {\em finite} number of additional bonds
  between nodes $y$ and $z$ with $d(y,z) < \infty$ doesn't change the
  internal scaling dimension either. Therefore we can slightly
  generalize lemma \ref{lem4} by changing our requirements to the
  following. Only bonds between nodes of finite distance and
  only finitely many bonds between nodes of distance $d(y,z) > k$ are
  inserted into \G ~to form \Gp. Then \Gp ~still has the same
  internal scaling dimensions $\underline{D}_S$ and $\overline{D}_S$
  as \G.
\end{remark}
\paragraph{Conclusions.}
We have seen that the internal scaling dimension does not depend on
the node from which we start our calculation and that under not too
strong conditions even the convergence of a subsequence of the relevant
sequence $D_n(x)$ is sufficient to calculate $\underline{D}_S$ and 
$\overline{D}_S$.
Furthermore the dimension is stable under local changes in the wiring
of the graph. This is a very desirable feature for physical reasons.
Furthermore it shows that a mechanism inducing dimensional phase 
transitions has to relate nodes of increasing distance, i.e.
has to change the graph non-locally. 

\subsection{Relations Between Internal Scaling Dimension and Connectivity
Dimension} \label{sec3.2}
As already stated above the two concepts of dimension we introduced are
not equivalent. In the following lemma we show that the existence of the
connectivity dimension implies the existence of the internal scaling
dimension and that they then have the same value.
\begin{lemma} \label{lem1}
  Let $x \in N$ again be an arbitrary node in \G. In the case that the
  limit ~$\lim_{n \rightarrow \infty} \frac{\ln|\partial
  \U_n(x)|}{\ln(n)} =: D_C(x)-1$ exists with $D_C(x) > 1$, 
  \G ~has internal scaling
  dimension $D_S(x)=D_C(x)$ starting from $x$.
\end{lemma}
\begin{proof}
The rather lengthy proof is given in detail in \cite{1a}. The main idea
is that $|\U_n(x)| = \sum_{j=0}^n |\partial \U_j (x)|$ such that
$|\U_n(x)|$ can be approximated with the knowledge of the behavior of
$|\partial \U_n(x)|$. After some calculations one gets the desired
result from this.
\end{proof}
Inversely, the existence of the internal scaling dimension does not
imply the existence of the connectivity dimension. There are examples
in which the scaling dimension has a well defined value but the
connectivity dimension does not exist. Even neither the upper nor the
lower connectivity dimension need to coincide with the scaling
dimension. In \cite{1a} we gave an example in which $D_S = D,
\underline{D}_C= 0$ and $\overline{D}_C= D+1$.  

The only always valid assertion is $\overline{D}_C +1 \leq D_S(x)$.

\section{Construction of Graphs} \label{sec4}
In the following we want to show how to construct graphs of arbitrary
real internal scaling dimension. We also want to investigate the
connections between the internal scaling dimension of graphs and the
box counting dimension of fractal sets. As will been seen below there
is a strong relationship between self similar sets and what we also
want to call self similar graphs with non-integer internal scaling
dimension.

\subsection{Conical Graphs with Arbitrary Dimension} \label{sec4.1}
For the sake of simplicity we concentrate our discussion on graphs
with dimension $1 \leq D \leq 2$. Graphs with higher dimension are
easily constructed using a nearly identical scheme.

\fig{Example of a $\frac{5}{3}$ dimensional conical graph}{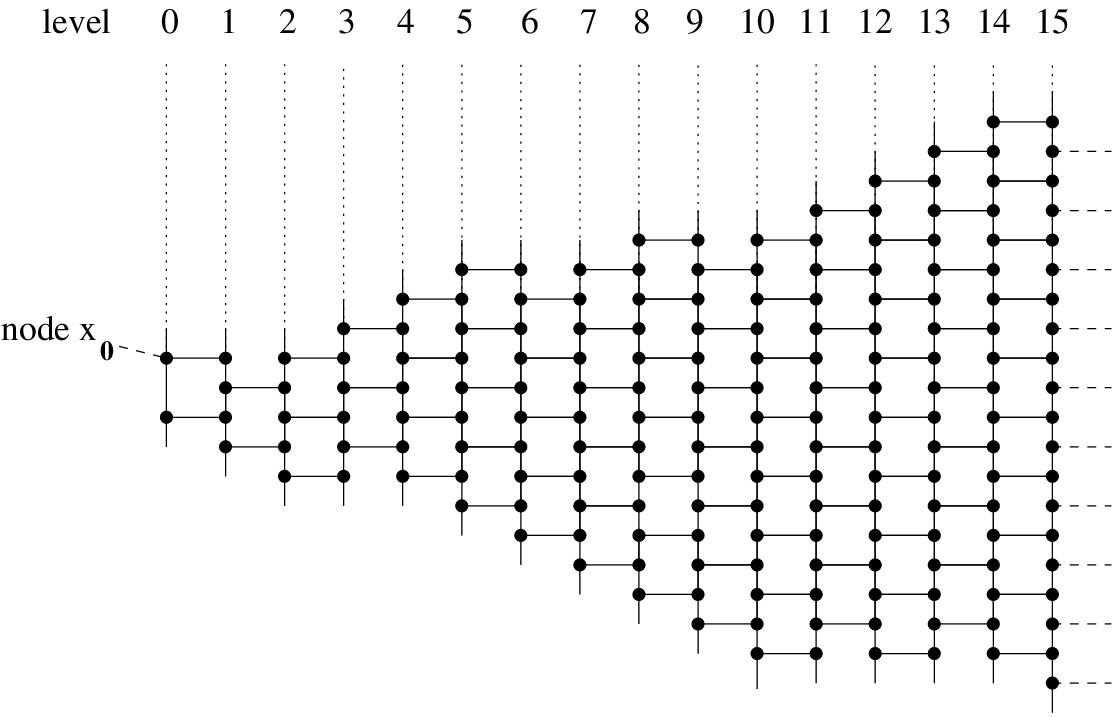}{tb}
Let $1 \leq D \leq 2$ be an arbitrary real number. Now we construct
the graph like in figure \ref{figbanana}. On level $m$ we use a height
of $\lfloor (2m - 1)^{D-1}\rfloor$ boxes. The construction is
continued ``to the right'' infinitely. To calculate the dimension we
observe that starting from $x_0$ we reach level $m$ after $n= 2m - 1$
steps. Thus we get with $n_k := 2k - 1$
\begin{align}
|\partial \U_{n_k}(x_0)| = \lfloor n_k^{D-1} \rfloor \thus \lim_{k
  \rightarrow  \infty} \frac{\ln|\partial \U_{n_k}(x_0)|}{\ln(n_k)}=
  D-1 \punkt
\end{align}
Using lemmas \ref{lem1}, \ref{lem2} and \ref{lem3} we see that this
graph has internal scaling dimension $D_S=D$.
If we close the construction vertically, i.e.\ introduce bonds between
the uppermost and the lowest nodes on each level we even can achieve
a completely homogeneous node degree $d=3$.

\begin{remark}
  Locally the constructed conical graph is completely isomorphic to
  a two-dimensional lattice. The non-integer dimension is
  only implemented as a global property of the graph.
\end{remark}

\subsection{Self-Similar Graphs} \label{sec4.2}
It is well known in graph theory that it is notoriously difficult to
construct large graphs with prescribed properties. It also proved
quite difficult to construct graphs with a prescribed (internal
scaling) dimension $D_S=D$ which don't exhibit the disadvantages of
the conical graphs described above. The main idea which solves the
problem is to use the well known theory of self similar sets or fractals
and their dimension theory. In the following we want to show
how this works and that we indeed can construct adjoint graphs to
self similar sets which have internal scaling dimension equal to the
box counting dimension of the self similar sets.

Given a strictly self similar set in $\R^p$ we canonically
construct an adjoint graph which also will be called
self-similar. 
The construction principle is based on an algorithm to compute the box
counting dimension of a self-similar set. 

For details concerning self-similar sets and dimensions of fractals
see \cite{6a}.

\subsubsection{Construction Based on Self-Similar Sets} \label{sec4.2.1}
Let $M$ be a strictly self-similar set with similarity transforms
$S_i$, $i \in I$, $I \subset \N$ and $|I| < \infty$. The contraction
factors $c_i$ of $S_i$ may all be equal, $c_i = c \in (0,1)$.
Now we cover $M$ with cubic lattices $L_n \subset \R^p$ with closed cubes of
edge length $c^n$, $n \in \N$, and replace every cube which has non-void
intersection with $M$ by a node. Nodes will be connected iff the
corresponding cubes in the covering cubic lattices have a non-void
intersection, i.e.~have a common corner or edge.

By this construction we get a finite graph $\G_n$ for
each $n \in \N$. The degree of these $\G_n$ is uniformly bounded
because an $n$-dimensional cube can only touch a finite number of
neighbor cubes in the cubic lattice. 
The graph we are interested in is $\G_\infty$, the graph we get through
infinite continuation of our construction. 
\begin{remark} \hfill
\begin{enumerate}
\item No problems arise from the infinite continuation of the
  construction steps.
\item The self-similarity of $M$ transfers to \G ~in the sense that we
  can also define an equivalent of the similarity transforms of the
  self-similar set $M$.
\item Connected self-similar sets produce connected self-similar
  graphs. The inverse is not true in general though.
\end{enumerate}
\end{remark}

\subsubsection{Self-Contained Construction Algorithm} \label{sec4.2.2}
There also are self-contained construction
algorithms for self-similar or hierarchical graphs.
One possibility is the following algorithm:
\begin{enumerate}
\item We start with a single node, $\G_0= (\{n_0\}, \emptyset)$.
\item $\G_1$ is the so-called generator, some finite graph.
\item We construct $\G_{n+1}$ from $\G_n$ by replacing every node in $\G_n$
by the generator $\G_1$ and interpret the original bonds in $\G_n$ as bonds
between some ``marginal'' nodes of the different copies of $\G_1$. 
\end{enumerate}
The construction is not unique. The result strongly depends on the
choice of the nodes in $\G_{n+1}$ which carry the bonds of $\G_n$.  In
our example all ``marginal'' nodes of the generator are equivalent
because of the symmetry of the generator and therefore the
construction is unique.

A slightly different construction algorithm with identical results is
also possible. We will not describe it here for lack of space. Refer
to \cite{1a} for more details.

\subsubsection{Dimension of Self-Similar Graphs} \label{sec4.2.3}
Now we calculate the dimension of the graphs we get by the above
construction using some self-similar set $M$. For the sake of simplicity 
we assume that $\G_1$ has a central node $x_0$ in the sense that all
``marginal'' nodes which carry the ``outer'' bonds have all the same 
distance $r$ to this node. We further assume that $\frac{1}{c}$ ($c$ the
contraction parameter) is a natural number which is true in
most of the well known examples of self-similar sets
 and finally that the self-similar set produces a connected adjoint graph.
Then it is easy to see that starting from node $x_0$ we can exactly reach 
all nodes of construction step $k+1$
after $n_{k+1}= r + 2r\,n_k + n_k = (2r+1) \,n_k + r$ steps
in the graph, with $n_0 = 0$.
\enlargethispage{0.5cm}
Thus $|\U_{n_k}(x_0)|$ is equal to the number of nodes in construction step
$k$, i.e. $|\U_{n_k}(x_0)| = N_{\delta_k} =
N_{c^k}$.\footnote{$N_{\delta_k}$ 
  is the number of cubes of edge length $\delta_k$ intersecting M,
  see the calculation of the box counting dimension in e.g. \cite{6a}.}
Explicitly we get for $n_k$
\begin{align}
n_k = \sum_{j=0}^{k-1} (2r+1)^{j} \,r = r \,\frac{(2r+1)^k - 1}{2r} \quad
\forall k \geq 1 \punkt
\end{align}
Now let us relate $r$
to the contraction parameter $c$ of the self-similar set. We assumed
that the graph constructed from the self-similar set is
connected. This implies that there are $\frac{1}{c}$ nodes on the
``diagonal'' of the generator, i.e. $2r + 1 = \frac{1}{c}$.
Now we have for the internal scaling dimension of \G
\begin{align}
\lim_{k \rightarrow \infty} D_{n_k}(x_0) & = \lim_{k \rightarrow \infty}
\frac{\ln(N_{c^k})}{\ln\left(r\frac{(2r+1)^k-1}{2r}\right)} \\ 
& =  \lim_{k \rightarrow \infty} \frac{\ln(N_{c^k})}{\ln((2r+1)^k) + \ln\left(
\frac{1-(2r+1)^{-k}}{2r}\right)} \\
& = \lim_{k \rightarrow \infty} \frac{\ln(N_{c^k})}{-\ln(c^k) +
\ln\left(\frac{1-(2r+1)^{-k}}{2r}\right)} = \dim_B(M) 
\end{align}
in which $\dim_B(M)$ is the box counting dimension of $M$.
Of course lemmas \ref{lem2} and \ref{lem3} provide us with the knowledge
that this is the dimension of \G ~starting from any node.

Thus we established equality of the box counting dimension of self-similar
sets and the internal scaling dimension of the adjoint self-similar 
graphs under the assumptions stated above.
\begin{remark}
The assumed existence of a central node $x_0$ is not essential for the
equality of the dimensions of the fractal and the graph. The equality 
still holds in a more general context, e.g. for fractals like the 
Sirpinski Triangle. It is difficult though to give a general proof for
arbitrary self-similar sets.
\end{remark}

\subsubsection{Approximation of a Two Dimensional Lattice} \label{sec4.2.4}

\fig{Some generators to approximate $\mathbb{Z}^2$}{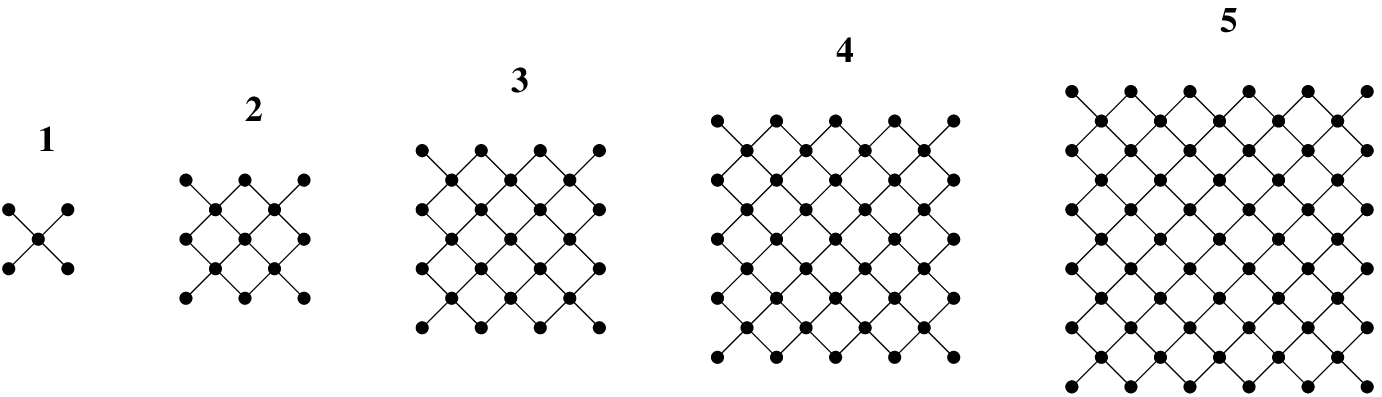}{tbp}
In this paragraph we want to show how it now becomes possible to do
a dimensional approximation of a $n$-dimensional cubic lattice.
Again, for the sake of simplicity, we discuss the idea only 
with a two-dimensional lattice but the generalization to $n$ dimensions 
is obvious.

We introduce generators as shown in figure \ref{figgener}. With these we
get graphs of dimensions
\begin{align}
D_S^{(l)} = \frac{\ln(2 l^2 + 2 l +1)}{\ln(2 l +1)} 
\end{align}
in which $l$ is the number which labels the generators in figure
\ref{figgener}.
Obviously we have
\begin{align}
\lim_{l \rightarrow \infty} D_S^{(l)} = \lim_{l \rightarrow \infty}
\frac{\ln(2 l^2 + 2 l +1)}{\ln(2 l +1)} = \lim_{l \rightarrow \infty}
\frac{2\ln(l)+\ln(2+\frac{2}{l} + \frac{1}{l^2})}{\ln(l) +
  \ln(2+\frac{1}{l})}
= 2 \punkt
\end{align}
In this sense we have a dimensional approximation of a
two-dimensional lattice as alleged above. 
This might have some relevance in
connection with the dimensional regularization used in many
renormalization approaches to quantum field theory.
\begin{remark}
The generators above correspond to fractal sets known as ``sponges'',
see e.g. \cite{7a}.
We can construct such ``sponges'' for any dimension $n$, we just need
to modify the generators appropriately. 
\end{remark}


\begin{thebibliography}{99}
\bibitem{1a}{T. Nowotny, M. Requardt, G\"ottingen preprint (to appear in
    J. Phys. A), hep-th/9707082}
\bibitem{6a}{K. J. Falconer, ``Fractal Geometry. Mathematical
    Foundations and Applications'', J. Wiley \& Sons, Chichester 1990}
\bibitem{7a}{G. A. Edgar, ``Measure, Topology and Fractal Geometry'',
    Springer, New York 1990}
\bibitem{Graham} R.L. Graham, D.E. Knuth, O. Patashnik, ``Concrete
    Mathematics'', Edison-Wesley, NY 2nd Edition 1988
\bibitem{1}C.Rovelli, ``Loop Quantum Gravity'', gr-qc/9710008
\bibitem{2}L.Smolin, ``The Future of Spin Networks'', gr-qc/9702030
\bibitem{3}C.J.Isham, ``Structural Issues in Quantum Gravity'',
  gr-qc/9510063 (Lecture given at the GR 14 conference, Florence 1995)
\bibitem{4}C.Castro, ``Beyond Strings\ldots'', hep-th/9707171
\bibitem{5}A.Connes, ``Non-commutative Geometry'', Acad.Pr. N.Y. 1994
\bibitem{6}Bollob\'as, ``Graph Theory'', Graduate Texts in Mathematics,
    Springer N.Y. 1979 
\bibitem{7}T.Toffoli, N.Margolus, ``Cellular Automaton Machines'', MIT
  Pr., Cambridge Massachusetts 1987
\bibitem{8}M.Requardt, ``Discrete Mathematics and Physics on the
  Planck-Scale\ldots'', hep-th/9605103
\bibitem{9}M.Requardt, ``A New Approach to Functional Analysis on
  Graphs\ldots'', hep-th/9708010
\bibitem{10}M.Requardt, ``Emergence of Space-Time on the Planck
  Scale\ldots'', hep-th/9610055
\bibitem{11}B.Bollob\'as, ``Random Graphs'', Academic Pr. London 1985
\bibitem{12}L.Nottale, ``Fractal Space Time and Microphysics\ldots'',
  World Scientific 1992
\bibitem{Di}A. Dimakis, F. M\"uller-Hoissen, J.Math.Phys {\bf 35}
    (1994) p. 6703
\bibitem{So} R.\ D.\ Sorkin, Int. J. Theor. Phys. {\bf 30} (1991) p. 923
\bibitem{Ba} A.P. Balachandran et al, J. Geom. Phys. {\bf 18} (1996) p. 163
\end{thebibliography}
\end{document}